\documentclass[american,english,aps,twocolumn,groupaddress,showpacs,prl]{revtex4}
\usepackage[T1]{fontenc}
\usepackage[latin9]{inputenc}
\usepackage{textcomp}
\usepackage{amstext}
\usepackage{graphicx}

\makeatletter

\newcommand*\LyXbar{\rule[0.585ex]{1.2em}{0.25pt}}

\@ifundefined{textcolor}{}
{%
 \definecolor{BLACK}{gray}{0}
 \definecolor{WHITE}{gray}{1}
 \definecolor{RED}{rgb}{1,0,0}
 \definecolor{GREEN}{rgb}{0,1,0}
 \definecolor{BLUE}{rgb}{0,0,1}
 \definecolor{CYAN}{cmyk}{1,0,0,0}
 \definecolor{MAGENTA}{cmyk}{0,1,0,0}
 \definecolor{YELLOW}{cmyk}{0,0,1,0}
 }

\usepackage{times}
\renewcommand{\citet}{\cite}
\usepackage{mathptmx}
\usepackage{graphicx}
\usepackage{bm}

\makeatother

\usepackage{babel}

\begin{document}

\title{Mass transportation of thermally driven nanotube nanomotors with
defects}

\author{Jige Chen$^{1}$}

\email{chenjige@sinap.ac.cn}

\author{Yi Gao$^{1}$}

\author{Chunlei Wang$^{1}$}

\author{Renliang Zhang$^{1}$}

\author{Hong Zhao$^{2}$}

\author{Haiping Fang$^{1}$}

\affiliation{$^{1}$Shanghai Institute of Applied Physics, Chinese Academy of
Sciences, Shanghai 201800, China}

\affiliation{$^{2}$Department of Physics, Xiamen University, Xiamen 361005, China}

\begin{abstract}
Thermally driven nanotube nanomotors provide linear mass transportation
controlled by a temperature gradient. However, the underlying mechanism
is still unclear where the mass transportation velocity in experiment
is much lower than that resulting from simulations. Considering that
defects are common in fabricated nanotubes, we use molecular dynamics
simulations to show that the mass transportation would be considerably
impeded by the potential barriers or wells induced by the defects,
which provides a possible picture to understand the relative low value
at microscopic level. The optimal structure and the factors which
would affect the performance are discussed. The result indicates considering
defects is helpful in designing nanotube nanomotor and other new nanomotor-based
devices.
\end{abstract}
\maketitle

\section{introduction}

Controlled mass transportation is the key function of the molecular
motor. Nature already provides some biological nanomotors, which however
can only work in specific environmental conditions\citet{01,02,03}.
In contrast, nanotube nanomotors\citet{04,05} can operate in diverse
environments that include various chemical media, as well as electric
or magnetic fields\citet{06,07,08,09,10,11,12}. Their multiple advantages
make them capably evolved into components of versatile nanodevices
in applications. Pressure gradients, mechanical force, and electrical
bias, et al. are the possible driving forces\citet{04,05,13,14} in
nanotube nanomotors. Recently, the use of thermal gradient to actuate
mass transportation have been demonstrated to be highly valuable in
nanotube nanomotor design\citet{08,09,10,11,12}. Thermophoresis,
also known as the Soret effect, is capable of driving fluids, gases,
DNA molecules and other nano materials that are subjected to a thermal
gradient. In 2008, the first successful fabrication of a thermally
driven nanotube nanomotor was reported by Barreiro et al\citet{09}.,
in which the outer tube of a double-walled carbon nanotube (DWNT)
traveled along a coaxial inner tube by actuation of a temperature
gradient. Later, mass transportation of carbon nanotube (CNT) capsules,
the inner tube of a DWNT, graphene nanoribbons and other nano materials
were experimentally realized or theoretically proposed\citet{08,10,11,15,16,17,18,19,20,21}.%
\begin{figure}
\includegraphics[scale=0.17]{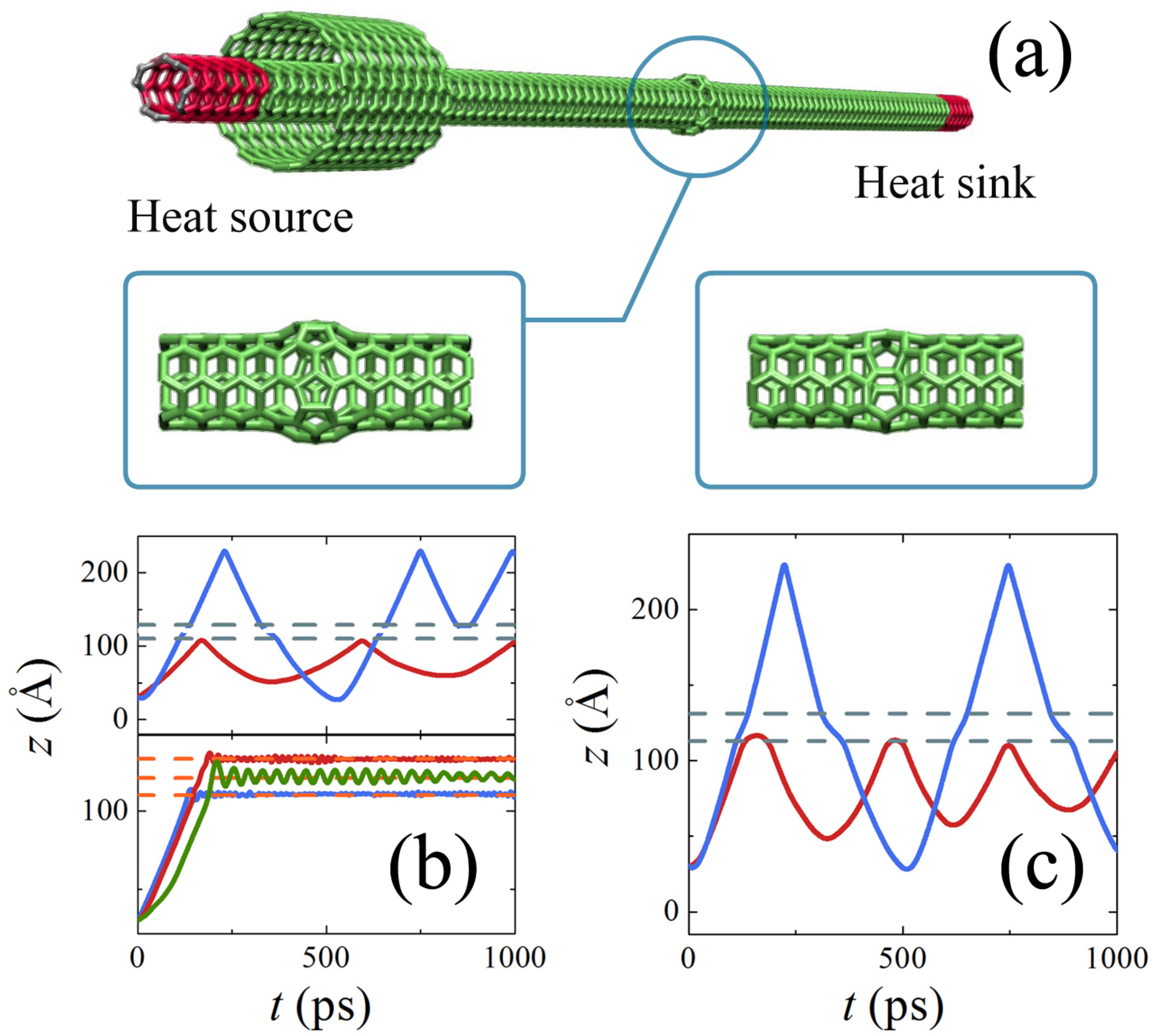}

\caption{System and kinetic results. (a) Schematic of the nanotube nanomotor.
The fixed atoms are silver, and the atoms at the heat source and heat
sink are red. The carbon ad-dimer (CD) defects (left) or Stone-Wales
(SW) defects (right) are placed in the middle of the inner tube. (b,
c) The axial trajectory z of the center of mass (COM) of the outer
tube as a function of simulation time, t, along the inner tube, with
(b) CD defects and (c) SW defects, as determined by independent simulations.
The dashed lines indicate the possible bouncing or trapping sites. }

\end{figure}
\begin{figure*}
\includegraphics[scale=0.18]{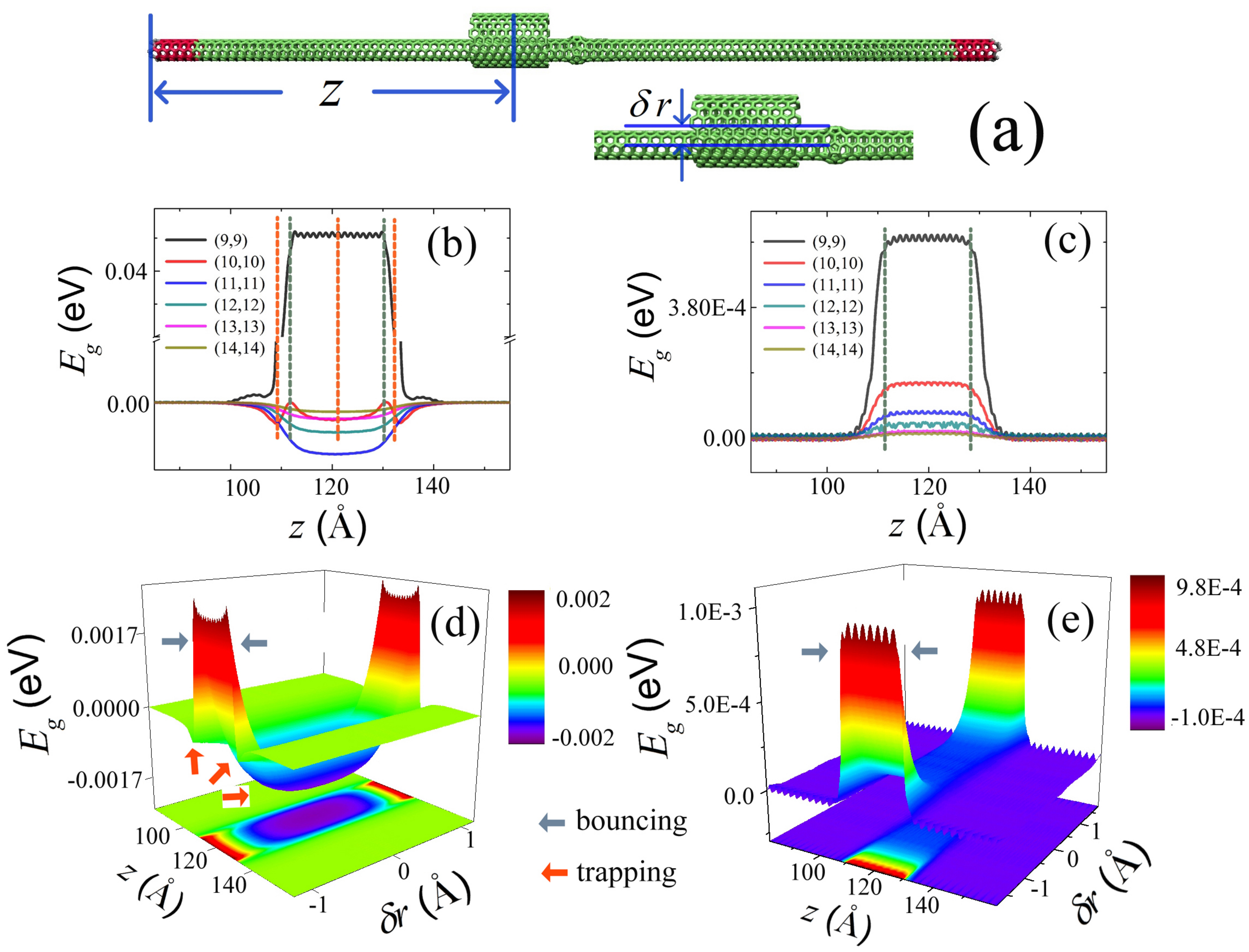}

\caption{Van de Waals energy change $E$$_{g}$ induced by defects. (a) Schematics
of the axial coordinate and the deviation distance $\delta r$. (b,
c) $E$$_{g}$ as a function of $z$ around the (b) CD defects and
(c) SW defects. The chirality vector of the outer tube ranges from
(9, 9) to (14, 14). The possible bouncing (trapping) sites are indicated
by dashed cyan (orange) lines. (d, e) $E_{g}$ as a function of both
$z$ and $\delta r$ around the (d) CD defects and (e) SW defects
in the (4, 4)/(11, 11) DWNT. The possible bouncing and trapping sites
are indicated by arrows.}

\end{figure*}

Although the underlying mechanism of thermophoresis is still unclear,
there is a growing interest in the scientific community to design
and fabricate thermally driven nanotube nanomotor due to its practical
usability and potential applications. We note that in most simulations,
the average mass transportation velocity is about 1-2 $\textrm{\AA}$/ps
(1-2\texttimes{}10$^{8}$ um/s), while it is only 1-2 um/s in experiments\LyXbar{}this
is 7 orders of magnitude lower than the simulation value. Besides
the small system dimension and large temperature gradient limited
by the calculation capabilities\citet{09}, no other picture is provided
to understand origin of the stagnation at microscopic level. Meanwhile,
defects are common in practically fabricated nanotubes according to
results from scanning tunneling microscope observations as well as
quantum and classical simulations\citet{22,23,24,25,26,27,28}. The
properties of CNTs would be drastically modified in the presence of
those defects. The most common defects, such as carbon ad-dimer (CD)
defects\citet{23,24,25} and Stone-Wales (SW) defects\citet{26,27,28},
are usually induced by one or more pentagon-heptagon (5-7) pairs in
CNTs. They produce changes in the topological structure and consequently
affect the electronic, mechanical, and thermal properties of CNTs.
Despite the significant impact and inevitable presence of defects,
their explicit effect upon the mass transportation of nanotube nanomotors
has not been reported. 

In this paper, we use molecular dynamics (MD) simulations to investigate
defective nanotube nanomotor and find out the mass transportation
might be considerably impeded by defects, which gives a possible picture
to understand the relatively low transportation velocity.

\section{Methods}

The nanotube nanomotor consists of a 24 nm long (4, 4) inner tube
and a 2 nm long outer tube with chirality vector ranging from (9,
9) to (14, 14). Two most common defects, namely, the carbon ad-dimer
(CD) and the Stone-Wales (SW) defects, are placed in the middle of
the inner tube. The CD defect is a 7-5-5-7 defect formed by adsorption
of a carbon dimer\citet{23,24,25}. The SW defect is a 5-7-7-5 defect
formed by the $\pi/2$ rotation of a C-C bond\citet{26,27,28}. Their
initial geometries are determined using a topological defects generating
algorithm based on the MM3 Allinger force field\citet{29,30}. Figure
1(a) shows the initial structure of the DWNT. An MD package LAMMPS\citet{31}
and the AIREBO potential\citet{32} are used to perform the MD calculations.
A minimum time step of 1 fs is employed for all of the simulations. 

The simulations are performed in three steps: 

(1) The first step consists of isothermal equilibration in which the
DWNT is thermalized at 300 K for 100 ps. 

(2) The second step consists of the non-equilibrium MD simulation.
Two slabs, one at each end of the inner tube, are used as the heat
source and heat sink. The temperature gradient is established by implementing
a constant heat flux (4 eV/ps) for 1 ns\citet{33}.

(3) The third step consists of the mass transportation of the outer
tube. At this stage, the restriction on the outer tube is removed.
It is the actual production run for 1 ns.

\section{Results and Discussion}

Figure 1 shows the typical axial trajectories of the center of mass
(COM) of the outer tube. When encountering CD defects, the outer tube
may exhibit three phenomena: (1) It passes through the defects, (2)
It bounces back, and (3) It is trapped at some specific sites. Similarly,
upon encountering the SW defects, the outer tube exhibits (1) and
(2) phenomena. The dash lines in Fig. 1 represent the possible bouncing
and trapping sites. Furthermore, the outer tube is still possible
to pass through the defects after bouncing back or trapping for a
long time.

To understand the microscopic mechanism of the impedance, we analyze
the change of van de Waals energy induced by the defects as: 

\begin{equation}
E_{g}=\frac{\sum_{i=1}^{N}V_{i}^{'}-V_{i}}{N}\end{equation}

\noindent where $V_{i}^{'}$ (\foreignlanguage{american}{$V_{i}$})
is the van de Waals energy between the $i$th atom in the outer tube
and the other atoms in the DWNT with defective (perfect) inner tube,
and $N$ is the total number of atoms in the outer tube. Therefore
$E_{g}$ describes the average change of van de Waals energy when
the outer tube encounters the defects. $E_{g}$ varies according to
the configuration between the inner and the outer tubes. For simplicity,
we only consider three factors: (1) Axial coordinate $z$ of the COM
of the outer tube, (2) Deviation distance $\delta r$ between the
two tubes in Fig. 2(a), (3) Deviation angle $\delta\alpha$ between
the two tubes in Fig. 4. Since $\delta\alpha$ is small during the
mass transportation process (see Fig. 4(d) later), its contribution
is usually neglected. We first investigate the van de Waals energy
change induced by the CD defects. Fig. 2(b) shows $E_{g}$ varying
as a function of $z$, and Fig. 2(d) shows $E_{g}$ varying as the
function of both $z$ and $\delta r$. It shows that the possible
bouncing sites correspond to the edges of the potential barriers,
and the possible trapping sites correspond to the bottoms of potential
wells. Similar $E_{g}$ distributions are observed around the SW defects
in Fig. 2(c) and (e).

The above observations indicate that at microscopic level, defects
ruins the perfect crystal structure and remarkably impede the mass
transportation of the nanotube nanomotor. It leads to a possible relationship
between the low transportation velocity and surface roughness at macroscopic
level. The characteristic of the defects are represented by the van
de Waals energy change $E_{g}$, rather than the absolute potential
energy $V_{i}$ or $V_{i}^{'}$. As shown in Fig. 2, the defects are
quite small comparing with either the inner or outer tube. Once away
from the defects, $E_{g}$ drops to approximately zero quickly. The
kinetic behavior change (bouncing and trapping) also occurs near the
defects. Therefore, in the large fabricated DWNT system for applications
(with either larger diameter or tube length) or considering other
nanotube system with different potential parameters, although the
potential energy $V_{i}$ or $V_{i}^{'}$ may differ from present
simulations, similar van de Waals energy change $E_{g}$ would lead
to similar kinetic behaviors. %
\begin{figure}
\includegraphics[scale=0.4]{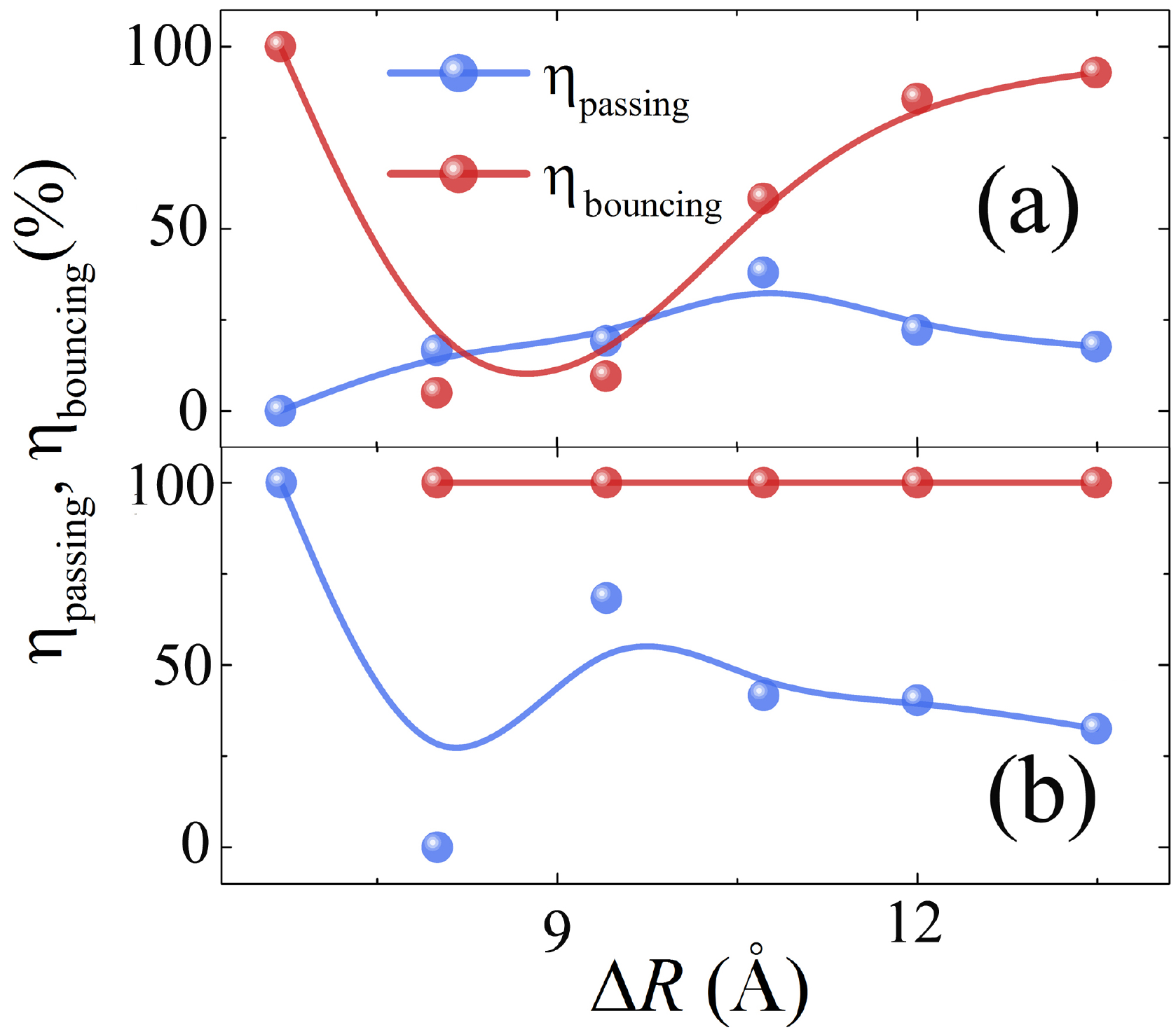}

\caption{(a, b) Passing ratio $\eta$$_{passing}$ and bouncing ratio $\eta_{bouncing}$
as a function of diameter difference $\Delta R$ for the (a) CD defects
and (b) SW defects. }

\end{figure}
\begin{figure}
\includegraphics[scale=0.12]{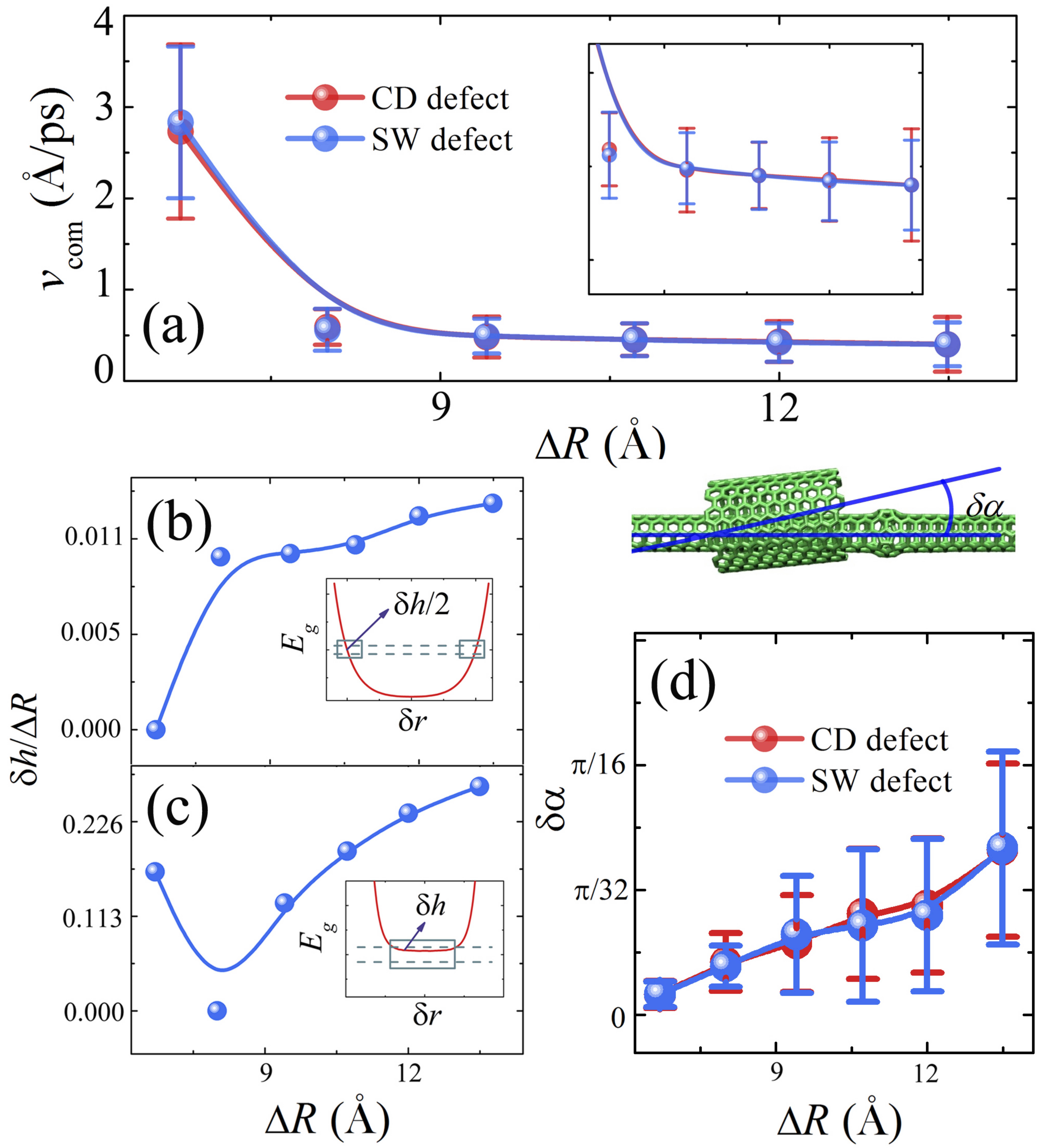}\caption{(a) The average velocity of the COM of the outer tube $v_{COM}$ as
functions of diameter difference $\Delta R$. (b, c) $\delta h/\Delta R$,
the configuration ratio of passing through the (b) CD defects and
(c) SW defects by only considering . $\delta h$ denotes the value
span of the proper $\delta r$ in which $E_{g}<E_{k}$, $E_{g}>0$
and $-E_{g}>-E_{k}$, $E_{g}<0$. (d) The average value of the deviation
angle $\delta\alpha$ as a function of $\Delta R$. }

\end{figure}

We attempted to determine the optimal structure to minimize the stagnation.
Therefore we define a passing ratio to measure the robustness of the
nanotube nanomotor as: 

\begin{equation}
\eta_{passing}=\frac{N_{passing}}{N_{passing}+N_{bouncing}+N_{trapping}}\end{equation}

\noindent where $N_{passing}$, $N_{bouncing}$, and $N_{trapping}$
are the numbers of times that the outer tube passes through the defects
in less than 30 ps, bounces back, and is trapped respectively. We
also define a bouncing ratio to measure how often the outer tube bounces
back in its failure of passing through as:

\begin{equation}
\eta_{bouncing}=\frac{N_{bouncing}}{N_{bouncing}+N_{trapping}}\end{equation}

\noindent In Fig. 3(a) and (b) we show $\eta_{passing}$ varying with
the diameter difference $\Delta R$ between the two tubes. Considering
the possibility that both CD and SW defects simultaneously exist,
we suggest that a proper diameter difference (neither too small nor
too large) is necessary to obtain the optimal passing ratio. Figure.
3 also shows $\eta_{bouncing}$ varies with $\Delta R$. For the CD
defects, $\eta_{bouncing}$ increases with $\Delta R$, which means
the failure of passing through is more due to the potential barriers
when enlarging the diameter difference. For the SW defects, $\eta_{bouncing}$
is always 100\% since only potential barriers are observed in the
associated $E_{g}$ distribution. %
\begin{figure}
\includegraphics[scale=0.45]{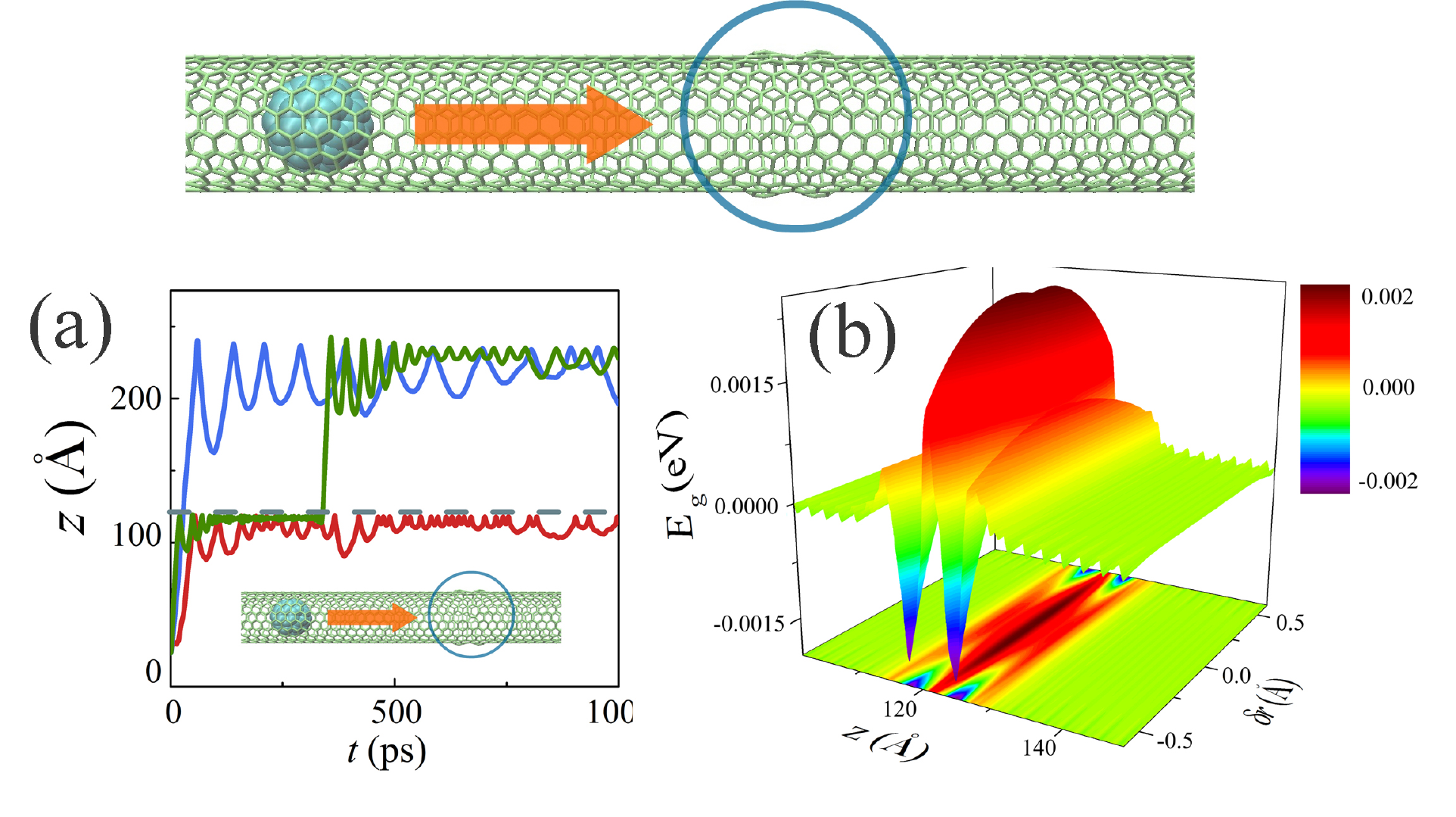}\caption{(a, b) Axial coordinates of a C$_{60}$ particle and its associated
van de Waals energy change $E_{g}$ around the SW defects. }

\end{figure}

Now we intend to understand the trend of the passing ratio by investigating
the possible configurations of the DWNTs. Fig. 4(a) shows the average
velocity of COM ($v_{COM}$ ) of the outer tube decreases with $\Delta R$,
and consequently the kinetic energy $E_{k}=\frac{1}{2}mv_{COM}^{2}$
also decrease with $\Delta R$. The deviation distance $\delta r$
varies from 0 to $\Delta R$ and only a proper $\delta r$ enables
the outer tube to pass through the defects, in which the kinetic energy
is larger than the potential barriers ($E_{g}<E_{k}$, $E_{g}>0$)
and potential wells ($-E_{g}>-E_{k}$, $E_{g}<0$). We denote $\delta h$
as the value span of the proper $\delta r$ which fulfills those requirements.
Therefore, the configuration ratio of passing through can be estimated
by $\delta h/\Delta R$. Fig. 4(b) and (c) show $\delta h/\Delta R$
takes a quite similar trend as the passing ratio in Fig. 3, which
explains the increase of $\eta_{passing}$ with $\Delta R$. However,
after reaching the maximum value, $\eta_{passing}$ decreases with
$\Delta R$. To explain it, we propose that the contribution of $\delta\alpha$
cannot be totally neglected in outer tube with large diameter, which
may result in the decrease of $\eta_{passing}$. In Figure. 4(d) we
show the average value of indeed increase with in the DWNTs.

Technically speaking, the mass transportation subject could be objects
inside the inner tube\citet{11,15,34,35,36}. Fig. 5(a) shows the
mass transportation of a fullerene (C$_{60}$) encapsulated in a (10,
10) CNT would be impeded by the defects with similar kinetic behaviors
as the nanotube nanomotor in Fig. 1. It also shows that the C$_{60}$
particle is still able to pass through the defects after a relative
long time. Fig. 5(b) illustrates the distribution with associated
potential barriers/wells. Adjusting to a proper configuration enables
the C$_{60}$ particle escape the defects.

\section{Conclusion}

In summary, we performed MD simulations on DWNTs to study the impact
of defects upon the mass transportation in a nanotube nanomotor. The
present simulation results have demonstrated that defects may considerably
impede the thermophoretic mass transportation by the associated potential
barriers and potential wells. It provides a possible picture to understand
the low transportation velocity in experiments at microscopic level.
Considering the impact of defects in fabricated nanotubes, we propose
that a proper choose of diameter difference is essential to achieve
the optimal robustness against defects. Our results will lead to improved
designs and applications of nanotube nanomotors in nanoengineering.
\begin{acknowledgments}
This work was supported by National Natural Science Foundation of
China (No. 21273268, 10925525, 11204341, 11290164), and Shanghai Committee
of Science and Technology under Grant No. 11DZ1500400. The authors
thank the High Performance Computing and Data Center, Shanghai Advanced
Research Institute, Chinese Academy of Sciences.\end{acknowledgments}

\end{document}